\documentclass{article}
\usepackage{multirow}
\usepackage[normalem]{ulem}
\usepackage{graphicx}
\usepackage{amsmath}
\usepackage{float}
\usepackage{placeins}
\useunder{\uline}{\ul}{}
\title{Dynamic investment model of the life cycle of a company under the influence of factors in a competitive environment.}
\date{}
\author{
O. A. Malafeyev, I. I. Pavlov \\ 
St.Petersburg State University, 7/9 Universitetskayanab., St.Petersburg,\\ 
199034 Russia}
\begin{document}
\maketitle
\begin{abstract}
Modelling all possible life cycles of a company in a highly competitive economic environment gives a significant advantage to the owner in his business investment activities. This article proposes and analyses a dynamic model of a company's life cycle with known action costs and transition probabilities, that can be affected by an outside influence. For this task, the Markov model was utilized. The proposed model is illustrated on a task of determining an advertising policy for a car dealership, that would increase the stock equity of a company. The result demonstrates the usefulness of a model for use in determining future actions of a company.  We also review multiple models of the influence of outside factors on a company's total capitalization.
\end{abstract}
\section{Introduction.}
Dynamic decision-making game is the basis for modelling the influence of positive and negative factors on a company's total capitalization. There are multiple possible models that can be applied to solve this problem. The Deterministic model takes into account the owner of a company, set of all possible states, discrete time variable, control function and some rational function. The Stochastic model adds to this a probability of transition from one state to another. We demonstrate stochastic model on an example by building a decision tree and finding an optimal strategy for the release of a new product on the national market. We also look at the probabilistic - deterministic model and demonstrate it on a simple example. 
 
Markov models are very effective in many different fields which rely on sequential decision making. At each step, the decision maker considers the current state and makes a decision with a goal to maximize a company's total capitalization, which leads a company to the next state. Action costs and transition probabilities are mainly responsible for how the system behaves.

As an example of an application of this model, we use a car dealership's marketing campaign. We apply our model to help the owner chose an optimal advertising policy for every city. This model uses transitional probabilities and profits as input data for Howard's improvement algorithm. This algorithm is chosen because of its simplicity and effectiveness in deciding which policy is optimal.

Many ideas that are used in this work are taken from [1]-[60]      
\section{Literature Review.}
There are some previously conducted studies related to the use of the Markov model which are summarized below.\\
Merve Merakli and Simge Kucukyavuz used Markov Decision Processes to solve an inventory management problem for humanitarian relief operations during a slow-onset disaster.\\
C. Drent, S. Kapodistria, and J. A. C. Resing formulated condition based maintenance policies under imperfect maintenance at scheduled and unscheduled opportunities as a semi-Markov decision process, and used it to find an optimal policy.\\
Yu-Yang Zhou and Guang Cheng used  Multi-Objective Markov Decision Processes in their novel shuffling method against DDoS attacks.\\
Sebastian Junges, Erika Abraham, Christian Hensel, Nils Jansen, Joost-Pieter Katoen, Tim Quatmann, Matthias Volk developed multiple analysis algorithms for parametric
Markov chains and Markov decision processes.\\
Thiago Freitas dos Santos, Paulo E. Santos, Leonardo A. Ferreira, Reinaldo A. C. Bianchi, Pedro Cabalar used Markov Decision Process to solve spatial puzzles.\\
Takashi Tanaka, Ehsan Nekouei, Ali Reza Pedram, Karl Henrik Johansson showed that mean-field approximation of a dynamic traffic routing game over an
urban road network leads to the linearly solvable Markov
decision process.

\section{Deterministic model of the influence of positive and negative factors on activities of a company's investment activities.}
\subsection{Informal formalization of the task of constructing and analysing a deterministic model of the influence of positive and negative factors on investment activities of a company.}
Let be $$\Gamma = \Gamma\langle N, X = \lbrace X_i^j \rbrace_{0_1}^{n^m}, t = \lbrace 0,1,...,n \rbrace , U = \lbrace u_i^j \rbrace_{1_1}^{n^m}, R = \lbrace r_i^j\rbrace_{1_1}^{b^m} \rangle$$ -- dynamic decision-making game by an owner of a business with full information. Here N - the owner of a company, making the decisions. X - set of all states of a company. t - discrete value, denoting time. U - control function, determining the transition from state $X_i$ to state $X_{i+1}, U_{i+1} : X_i \rightarrow X_{i+1}$. R - rational function is given on the set of all controls. The sum value of this function on controls is called a total win. Depending on what decision is made(control at the current stage) development of a company occurs according to one of the schemes, which implies an action of the collection, of positive factors $\alpha_i$, and negative factors $\beta_i$, while a company is in state $\Omega$. 
	
Among controls, emerging from the point X we are required to choose a control with maximum sum total capitalization.

\subsection{Formalization of the task of constructing and analysing a deterministic model of the influence of factors on a company's investment activities.}
Initially, the owner starts in a state $X_0$ and makes a management decision $u_1^j \in U_1$, i.e chooses one of the possible controls in this state. After that company transitions to state $X_1$. After that, internal and external factors with positive and negative influence start to affect company's investment activities. Because of this growth or decline of capitalization of a company is possible.

For example, the purchase of new equipment causes an increase in production volume. That, in turn, can cause multiple outcomes. The best one being -- an increase in profits.
With stable demand, purchasing new equipment with higher performance is inadvisable.

During the second step, the owner again makes a management decision $u_2^j \in U_2$, by choosing from a set of available controls. A company transitions to the next state - $X_2$.After that, internal and external factors with positive and negative influence start to affect a company, and so on.
  
On step n the owner chooses $u_n^j \in U_n$ one of the possible controls during this stage. After that a company transitions to the final step. During a transition from $X_i$ to $X_{i+1}$, a company's capitalization rises or declines. In final state $X_n$ capitalization of a company is defined as total capitalization during a previous step.

Thus, the problem will be solved, then a company's capitalization is be defined for each point and a maximum will be found.

\section{Stochastic model of the influence of negative and positive factors on a company's investment activities.}
\subsection{Informal formulation of the problem of building and analysing a stochastic model of the influence of negative and positive factors on a company's investment activities.}

Let be $$\Gamma = \Gamma\langle N, X = \lbrace X_i^j \rbrace_{0_1}^{n^m}, t = \lbrace 0,1,...,n \rbrace , U = \lbrace u_i^j \rbrace_{1_1}^{n^m}, P = \lbrace p_i^j \rbrace_{1_1}^{n^m}, R = \lbrace r_i^j\rbrace_{1_1}^{b^m} \rangle$$ -- a dynamic decision-making game for the owner of a company, where N -- is the owner, who makes decisions. X -- a set of all possible states of a company. t -- a discrete value, denoting time, which determines the step number. U -- a control function, which defines a transition from state $X_i$ to state $X_{i+1}$. R -- a rational function, which is defined on a set of all controls. A sum of function R is called  a sum total of company's capitalization. $p_i^j$ -- a probability of transition from state $X_i$ to state $X_{i+1}$.

Among controls, emerging from the point X we are required to choose a control with maximum sum total capitalization.

\subsection{Formulation of the problem of building and analysing the deterministic model of influences by factors on a company's investment activities.}

The model is described as follows. On initial stage entrepreneur - is an owner of a company, in state $X_0$ makes a decision. Then, a company's investment activities are affected by both external and internal factors that have either positive or negative effects. Entrepreneur possesses information, that events of the second stage from $X_2$ may occur with a certain probability P.
The sum of even probabilities from each stage = 1. The entrepreneur knows possible outcome options, their probabilities and will there be a rise or fall of a company's capitalization, depending on his decision, and by what amount. The uncertainty lies in the fact that the entrepreneur cannot determine what influence external factors will have on the choice of a particular decision. 

\textbf{Example}.

A company is deciding on the introduction of a new product on the national market. Uncertainty lies in, how will the market react to a new product. Test launch of a new product in the regional market is being considered. This way, the initial decision, which a company needs to make is -- should product's marketing start on a regional level. A company states, that launch on a regional market requires an initial investment of four million dollars, and launch on the national market required an investment of eighty million dollars. If trial sales at the regional level are not conducted, the decision to enter the national market can be made immediately.

A company considers sales number as successful, mediocre or negative, depending on volume sold. For the regional level, this equates to sale volumes of 500, 200 and 50 thousand units, and for the national 10000, 5000 and 1000 thousand units respectively. Based on the data, from the regional testing of similar products, a company assesses the probabilities of these three outcomes as 0.2, 0.7 and 0.1. Besides this, after researching the data about the relation of regional sales results with subsequent sales on the national market, a company has managed to estimate the following conditional probabilities. 

\begin{center}
\begin{table}[h]
\begin{tabular}{|c|l|c|l|l|l|}
\hline
\multicolumn{2}{|c|}{\multirow{2}{*}{}}                                                                      & \multicolumn{4}{c|}{\begin{tabular}[c]{@{}c@{}}Conditional probabilities of sales\\ in the nationwide market\end{tabular}} \\ \cline{3-6} 
\multicolumn{2}{|c|}{}                                                                                       &                                & Successful                    & Mediocre                    & Negative                    \\ \hline
\multirow{3}{*}{\begin{tabular}[c]{@{}c@{}}Probability of sales\\ in the regional market\end{tabular}} & 0.2 & Successful                     & 0.75                          & 0.2                         & 0.05                        \\ \cline{2-6} 
                                                                                                       & 0.7 & Mediocre                       & 0.35                          & 0.5                         & 0.15                        \\ \cline{2-6} 
                                                                                                       & 0.1 & Negative                       & 0.05                          & 0.3                         & 0.65                        \\ \hline
\end{tabular}
\end{table}
\end{center}

Besides that, it is known, that each sale brings a profit of 25 dollars on both local and national markets. 

The problem is to make a valid strategy for entering (or not entering) the market with a new product.

Solution: We consider three main components for situations of this type. After that, we will build a decision tree for this problem.

Let us calculate decision tree's characteristics.

Probabilities of sales levels in the nationwide market without initial regional market launch:
\begin{itemize}
	\item Successful: $(0.75*0.2) + (0.35*0.7) + (0.05*0.1) = 0.4$
	\item Mediocre: $(0.2*0.2) + (0.5*0.7) + (0.3*0.1) = 0.42$
	\item Negative: $(0.05*0.2) + (0.15*0.7) + (0.65*0.1) = 0.18$
\end{itemize} 
Revenue from a national market launch:
\begin{itemize}
	\item Successful: $10000*25 = 250000$
	\item Mediocre: $5000*25 = 125000$
	\item Negative: $1000*25 = 25000$
\end{itemize}
Profits from a national market launch without initial regional market launch:
\begin{itemize}
	\item Successful: $250000 - 80000 = 170000$
	\item Mediocre: $125000 - 80000 = 45000$
	\item Negative: $25000 - 80000 = -55000$
\end{itemize}
Results on a national market without initial regional market launch: \\$0.4 * 250000 + 0.42 * 125000 + 0.18 * 25000 - 80000 = 77000$\\
\\
Revenue from a regional market launch:
\begin{itemize}
	\item Successful: $500 * 25 = 12500$
	\item Mediocre: $200 * 25 = 5000$
	\item Negative: $50 * 25 = 1250$
\end{itemize}
Profits from a regional market launch:
\begin{itemize}
	\item Successful: $12500 - 4000 = 8500$
	\item Mediocre: $5000 - 4000 = 1000$
	\item Negative: $1250 - 4000 = -2750$
\end{itemize}
Profits from a national market launch with initial successful regional market launch:
\begin{itemize}
	\item Successful: $250000 - 80000 + 8500 = 178500$
	\item Mediocre: $125000 - 80000 + 8500 = 53500$
	\item Negative: $25000 - 80000 + 8500 = -46500$
\end{itemize}
Probabilities of sales levels in nationwide market with initial successful regional market launch:
\begin{itemize}
	\item Successful: $0.2 * 0.75 = 0.15$
	\item Mediocre: $0.2 * 0.2 = 0.04$
	\item Negative: $0.2 * 0.05 = 0.01$
\end{itemize}
Result on a national market with initial successful regional market launch:\\ $250000 * 0.75 + 125000 * 0.2 + 25000 * 0.05 - 80000 + 12500 - 4000 = 142250$\\
\\
Profits from a national market launch with initial mediocre regional market launch:
\begin{itemize}
	\item Successful: $250000 - 80000 + 1000 = 171000$
	\item Mediocre: $125000 - 80000 + 1000 = 46000$
	\item Negative: $25000 - 80000 + 1000 = -54000$
\end{itemize}
Probabilities of sales levels in nationwide market with initial mediocre regional market launch:
\begin{itemize}
	\item Successful: $0.7 * 0.35 = 0.245$
	\item Mediocre: $0.7 * 0.5 = 0.35$
	\item Negative: $0.7 * 0.15 = 0.105$
\end{itemize}
Result on a national market with initial mediocre regional market launch:\\ $250000 * 0.35 + 125000 * 0.5 + 25000 * 0.15 - 80000 + 5000 - 4000 = 74750$\\
\\
Profits from a national market launch with initial negative regional market launch:
\begin{itemize}
	\item Successful: $250000 - 80000 - 2750 = 167250$
	\item Mediocre: $125000 - 80000 - 2750 = 42250$
	\item Negative: $25000 - 80000 - 2750 = -57750$
\end{itemize}
Probabilities of sales levels in nationwide market with initial negative regional market launch:
\begin{itemize}
	\item Successful: $0.1 * 0.05 = 0.005$
	\item Mediocre: $0.1 * 0.3 = 0.03$
	\item Negative: $0.1 * 0.65 = 0.065$
\end{itemize}
Result on a national market with initial negative regional market launch:\\ $250000 * 0.05 + 125000 * 0.3 + 25000 * 0.65 - 80000 + 1250 - 4000 = -12900$\\

\begin{figure}[h]
  \includegraphics[width=\linewidth]{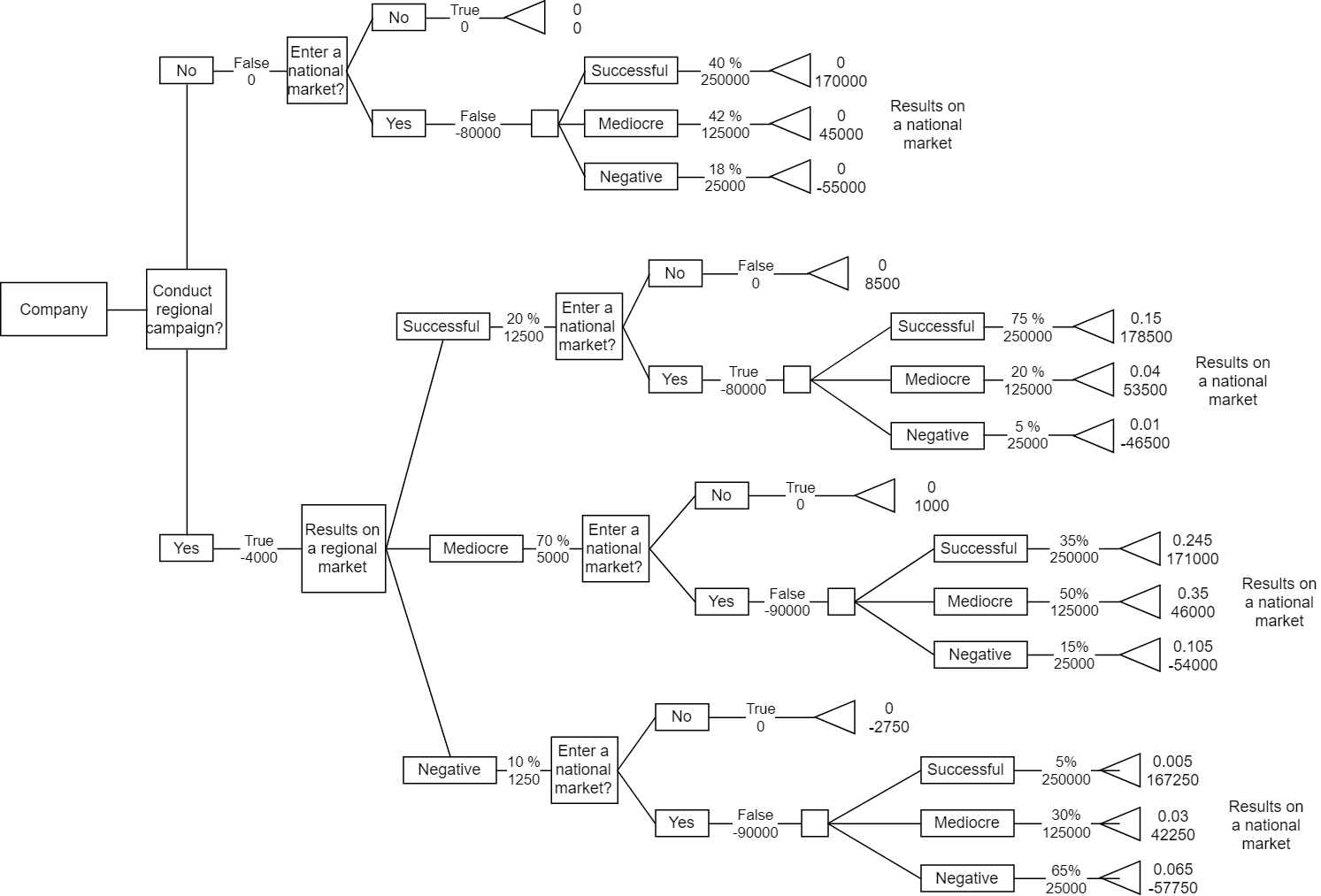}
  \caption{Product launch decision tree.}
  \label{fig:tree1}
\end{figure}

Thus, by analysing the decision tree, we can formulate an optimal strategy for a company, namely: it is necessary to make a preliminary sales in the regional market and to expand sales at the national level, only if results during regional trials prove to be successful.\\
\newpage

\section{ Probabilistic - deterministic model of influences by negative and positive factors on a company's investment activities.}
\subsection{Informal statement of the problem of constructing and analysing a probabilistic-deterministic model of influences by negative and positive factors on a company's investment activities.}
Let be $$\Gamma = \Gamma\langle N, X = \lbrace X_i^j \rbrace_{1}^{n}, t = \lbrace 0,1,...,n \rbrace , U = \lbrace u_t \rbrace_{1}^{n}, R = \lbrace r_k\rbrace_{1}^{c} \rangle$$- a dynamic game of decision-making by the owner of a company with complete information, where N is the owner of a company making decisions. X is the set of all company's states. t - discrete value denoting time. U - control function that determines the transition from state X to state $X_{i+1}, U_{i+1}:X_i \rightarrow X_{i+1}$. R - a rational function defined on the set of all controls.

\subsection{Formulation of the problem of building and analysing a probabilistic - deterministic model of influences by negative and positive factors on a company's investment activities.}
Now suppose that the choice of control at a point X does not determine a state, but only a probability distribution for this state. 

Let $X_t$ and $U_t$ be arbitrary finite sets. Each u from $U_t$ is mapped to probability distribution p on $X_t$. The function p, that defines the law of transition from $U_t$ to $X_t$, will be called the transition function. It is assumed that the point $X_0$ from the set from which the game begins is also random, and its probability distribution $\mu$              (initial distribution) is given.

The transition from x$\in X_{t-1}$ to $U_t$ is determined by the entrepreneur. In this case, we choose u not from all of $U_t$, but from its subset $U\left( x \right)$, that is dependant on the state x. The elements of a set $U\left( x \right)$ will be called controls at a point x. Sets $U\left( x \right)$ are defined and not empty for all non-final states. It is assumed that $U\left( x \right)$ do not intersect in pairs and their sum of all $x \in X_{t-1}$ is equal to $U_t$. In other words, each control u can be used in one and only one state. We denote this state as $j\left( u \right)$, so the record $x = j\left( u \right)$ is equivalent to the record $u \in U\left( x \right)$.

On the set of all controls, the current growth or decline of a company's capitalization $r\left( u \right)$ is given, and on the set of final states - the final growth or decline of a company's capitalization $r\left( x \right)$ is given.

The model is described as follows. At the initial moment of time, the entrepreneur - the owner of a company - makes a decision, chooses a control - a set of transition probabilities at the current stage. Then the second participant - nature chooses the next state. For the transition from state 0 to state 1, the owner receives a certain capitalization of a company.

The purpose of the model is to obtain the maximum capitalization of a company.

\textbf{Example}.

The column $U_1$ shows three probability distributions on the set $X_1$,
corresponding to the three controls from $X_0$. The column $U_2$ indicates five probability distributions on the set $X_2$, that correspond to the controls from the state $X_1$. The problem will be solved if we find the control with the maximum increase in the capitalization of a company. 

In state 3 expected value equals:
\begin{center}
$0 + \frac{1}{2}*1 + \frac{1}{2}*2 = \frac{3}{2}$
\end{center}
when choosing the first control and equals 
\begin{center}
$1 + \frac{1}{4}*1 + \frac{3}{4}*2 = \frac{11}{4}$
\end{center}
when choosing the second. The estimate of state 3 is equal to the maximum of these two numbers, i.e. $\frac{11}{4}$ and it is clear that under state 3 one should prefer the second control.
\newpage
\begin{figure}[H]
  \includegraphics[width=\linewidth]{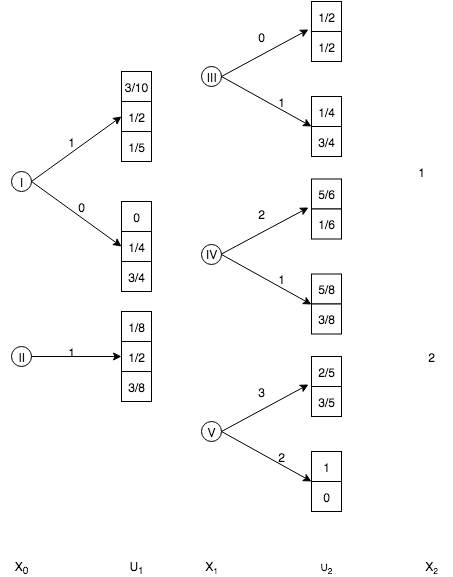}
  \caption{Model.}
  \label{fig:boxes1}
\end{figure}

Similarly,

\begin{center}
$v \left( IV \right) = max \left( 2 + \frac{5}{6}*1 + \frac{1}{6}*2; 1 + \frac{5}{8}*1 +\frac{3}{8}*2\right) = \frac{19}{6}, $
\end{center}

in state 4, the first control is preferable.
\begin{center}
$v \left( V \right) = max \left( 3 + \frac{2}{5}*1 + \frac{3}{5}*2; 2 + 1*1 + 0*2\right) = \frac{23}{5}, $
\end{center}

In state 5, the first control is more profitable. Next, choosing in state 1 the first control, and then, acting in an optimal way, we obtain an estimate,
\begin{center}
$1 + \frac{3}{10} * \frac{11}{4} + \frac{1}{2} * \frac{19}{6} + \frac{1}{5} * \frac{23}{5} = \frac{2597}{600}$
\end{center}

and choosing the second state, we obtain an estimate
\begin{center}
$0 + 0 * \frac{11}{4} + \frac{1}{4} * \frac{19}{6} + \frac{3}{8} * \frac{23}{5} = \frac{151}{60}$
\end{center}

\section{Stochastic decision-making investment model of company's life cycle.}
\subsection{Informal statement of the problem of constructing and analysing a stochastic decision-making investment model of company's life cycle.}
A model $\Gamma = \langle I = \lbrace 1,2,...,n \rbrace = \lbrace i \rbrace_1^k, K_i = \lbrace 1,2,...,k_i \rbrace , S_i = \lbrace s_i \rbrace_1^n , R_i \rangle$ is considered, in which the decision-making process is represented by a finite number of states i, i = 1,2, ..., n.

Let each state $i \in S = \lbrace 1,...,N \rbrace$ correspond to a finite set of solutions (or alternatives) $K_i$, the elements of which we denote by numbers 1,...,$k_i$. The space of strategies K is the direct product of sets of solutions K = $K_1 \times K_2 \times... \times K_n$.

Exists a matrix $P = (p_ij)$ of transition probabilities between states. The reward structure is represented in the form of a matrix $R \left( q \right) = \left( r_{ij} \left( q \right) \right)$, the elements of which are the income values. The matrix of transition probabilities P and the income matrix $R \left( q \right)$ depend on the decision policies k by the decision maker.

Depending on what decision is made (control at the current stage), the development of a company proceeds according to one or another scheme, which implies an influence by a combination of positive factors $\alpha_i$ and negative factors $\beta_i$ on the state of a company $\Omega$.

There is a finite set of company's states i, which is estimated by the value of a company's capitalization per unit of time t. Depending on the decision taken, a company goes into one of the possible states i. In each state i, the set of rational strategies k depends on this state.

The goal of the task is to find the optimal strategy that maximizes the expected capitalization of a company from a process that has a finite number of stages.

\subsection{Formulation of the problem of constructing and analysing a stochastic decision making investment model of company's life cycle.}
Let the process of transition from state to state occur, not deterministically, but stochastically and controlled by the transition matrix $P = \left( p_{ij} \right)$, where $p_{ij}$ is the probability that the system at the time t+1 is in the state j, if it is known, that at the time t it was in the state i.

Consider the case, when the matrix does not depend on time, and decisions are made at each step. Suppose that at each step one of the sets of such matrices can be chosen as a transition matrix, and we denote the matrix, corresponding to policy q, as $P \left( q \right) = \left( p_{ij} \left( q \right) \right)$.

Suppose further that not only the state at each step changes but also the capitalization of a company, which is a function of the initial and final states and the solution.

Let $R \left( q \right) = \left( r_{ij} \left( q \right) \right)$ denote the corresponding matrix of a company's capitalization.

Note that this model is not perfect since the same factor can be both positive and negative. Examples are a change in technology, personnel, investment, etc. In order to get the closer details of specific tasks, the stochastic model will be discussed below using a specific example.

\section{Classification of a company's investment activities in the Markov model of control of its activities.}
We introduce a classification based on a company's key performance indicators (for example, sales of goods and/or services, productivity, profits, customer satisfaction, etc.); in this case, we will consider the growth of stock prices, x(t). We introduce the following indicators:
\begin{itemize}
	\item growth rate, $V \left( t \right)  = \frac{x \left( t \right)}{t}$;
	\item growth acceleration, $a \left( t \right)  = \frac{V \left( t \right)}{t}$.
\end{itemize}

Thus, if the growth rate is zero, a company is in a stable state (there is no rise and fall in stock prices); if the growth rate is greater than zero, then a company is in a state of growth, because its shares are growing; if it is less than zero, then a company is in decline (stock quote falls). Acceleration largely characterizes stable conditions not only of a company but also of the stock market as a whole, since It shows spasmodic processes of rising or falling stock prices.

\subsection{Car dealership}

Suppose that a car dealership operates in five cities. In any of these cities we can invest in an advertisement in the following areas:
\begin{itemize}
	\item 1 - advertisement on the radio;
	\item 2 - placing ads on the website;
	\item 3 - placement of advertisements in the printed publication "Car Bulletin";
	\item 4 - TV adverts;
	\item 5 - active search for customers.
\end{itemize}

In each city for all of these fixed policies, the probabilities are set that the next customer will acquire a car in one of the five cities and the corresponding income in monetary terms associated with each sale. Since each policy has a different set of clients, transition probabilities and revenues depend on the policies. Income, in this case, is expressed as a fall or an increase in the capitalization of a company.

If we assign some hypothetical numbers to them, then these tasks can be written in a table.

% \usepackage{multirow}
\begin{table}[h]
\begin{tabular}{|l|l|l|l|l|l|l|l|l|l|l|l|}
\hline
\multirow{2}{*}{\begin{tabular}[c]{@{}l@{}}State\\ (city)\end{tabular}} & \multirow{2}{*}{Policy} & \multicolumn{5}{l|}{Transition probability} & \multicolumn{5}{c|}{Profit}       \\ \cline{3-12} 
                                                                        &                         & j = 1    & 2      & 3      & 4      & 5     & i = 1 & 2    & 3    & 4    & 5    \\ \hline
1                                                                       & 1                       & 0.1      & 0.2    & 0.4    & 0.1    & 0.2   & -300  & -200 & -190 & 280  & -120 \\ \hline
                                                                        & 2                       & 0.2      & 0.2    & 0.1    & 0.5    & 0     & 940   & -270 & -600 & 80   & 620  \\ \hline
                                                                        & 3                       & 0.7      & 0.1    & 0      & 0      & 0.2   & -60   & 100  & 520  & -760 & 230  \\ \hline
                                                                        & 4                       & 0.2      & 0.1    & 0.2    & 0.3    & 0.2   & 20    & 110  & 0    & 390  & 510  \\ \hline
                                                                        & 5                       & 0.2      & 0.1    & 0.1    & 0.4    & 0.2   & 620   & 0    & -640 & 250  & 630  \\ \hline
2                                                                       & 1                       & 0.1      & 0.2    & 0.2    & 0.2    & 0.3   & 0     & 330  & 680  & -240 & 390  \\ \hline
                                                                        & 2                       & 0        & 0.1    & 0.1    & 0.3    & 0.5   & 270   & 640  & 320  & 0    & -480 \\ \hline
                                                                        & 3                       & 0.1      & 0.5    & 0.2    & 0      & 0.2   & 460   & -480 & 0    & 180  & 570  \\ \hline
                                                                        & 4                       & 0.3      & 0.2    & 0.3    & 0      & 0.2   & -320  & 840  & 750  & 730  & 690  \\ \hline
                                                                        & 5                       & 0.4      & 0.3    & 0.1    & 0.2    & 0     & 560   & 510  & -90  & 120  & 0    \\ \hline
3                                                                       & 1                       & 0.3      & 0      & 0.4    & 0.2    & 0.1   & 170   & 120  & 180  & -300 & 280  \\ \hline
                                                                        & 2                       & 0.5      & 0.2    & 0      & 0.2    & 0.1   & 220   & -60  & 150  & 0    & 660  \\ \hline
                                                                        & 3                       & 0.1      & 0.2    & 0.5    & 0.1    & 0.1   & 110   & 20   & 100  & -160 & 90   \\ \hline
                                                                        & 4                       & 0        & 0.1    & 0.4    & 0.3    & 0.1   & 170   & 170  & 0    & 300  & -880 \\ \hline
                                                                        & 5                       & 0.4      & 0      & 0.1    & 0.4    & 0.1   & -660  & 480  & 810  & 830  & 240  \\ \hline
4                                                                       & 1                       & 0.2      & 0.2    & 0.1    & 0.1    & 0.4   & -100  & 0    & 800  & 150  & 560  \\ \hline
                                                                        & 2                       & 0        & 0.1    & 0.2    & 0.6    & 0.1   & 620   & 560  & 470  & -620 & 400  \\ \hline
                                                                        & 3                       & 0.5      & 0.1    & 0.1    & 0.2    & 0.1   & 170   & 250  & 0    & -280 & 0    \\ \hline
                                                                        & 4                       & 0.3      & 0.1    & 0.1    & 0.5    & 0     & 500   & 630  & 920  & 430  & 90   \\ \hline
                                                                        & 5                       & 0.3      & 0.3    & 0.1    & 0.1    & 0.1   & 80    & 510  & -110 & 110  & 300  \\ \hline
5                                                                       & 1                       & 0.1      & 0.3    & 0.1    & 0.4    & 0.1   & 710   & -160 & 50   & 280  & 690  \\ \hline
                                                                        & 2                       & 0.2      & 0.2    & 0      & 0.2    & 0.4   & 920   & 100  & 600  & 270  & 180  \\ \hline
                                                                        & 3                       & 0        & 0.1    & 0.1    & 0.7    & 0.1   & 0     & 740  & 710  & -60  & 160  \\ \hline
                                                                        & 4                       & 0.1      & 0.1    & 0.1    & 0.3    & 0.4   & 500   & 980  & 900  & 220  & 580  \\ \hline
                                                                        & 5                       & 0.4      & 0.1    & 0.3    & 0      & 0.2   & 520   & 400  & 960  & 60   & 690  \\ \hline

\end{tabular}
\end{table}

Lets us explain the table using the last row as an example. It shows that investment in the first policy (advertising on the radio) in the fifth city, will lead to a probability of 0.1 that there will be a sale in city 1, and there will be an increase in the capitalization of a company by 710 units. With a probability of 0.3, there will be a sale in city 2 with a loss of  160 units, and a probability of 0.1 that there will be a sale in city 3, with an income of 50 units, a probability of 0.4 that the sale will occur in city 4, with an income of 280 units and a probability of 0.1 that there will be a sale in city 5, and the capitalization of a company will increase by 690 units.

There are five states in this problem, i.e. N = 5, and five policies in states 1, 2, 3, 4, 5, i.e. $n_1= 5, n_2= 5, n_3= 5, n_4=5, n_5=5$. This means that there are 5 * 5 * 5 * 5 * 5 = 3125 possible policies
\\
\\
\\
Solution:

As the initial approximation, the policy vector D will be chosen as, 

\begin{center}
$D = \begin{bmatrix}1\\1\\1\\1\\1\end{bmatrix}$
\end{center}
which means that we work in all cities. This is a policy that maximizes the expected value of a company's capitalization growth. For this policy, we have a matrix of transition probabilities P 
\begin{center}
$P = \begin{bmatrix}0.1&0.2&0.4&0.1&0.2\\0.1&0.2&0.2&0.2&0.3\\0.3&0&0.4&0.2&0.1\\0.2&0.2&0.1&0.1&0.4\\0.1&0.3&0.1&0.4&0.1\end{bmatrix}$
\end{center}

The sum $\sum_{j=1}^N p_{ij}r_{ij}$ will be denoted as $q_i$, since the expected capitalization of a company depends only on i. The equation for determining the values of $v_i$, assuming a value of $v_3$ is zero, is written as
\begin{center}
$\begin{array}{lcl}
g + v_1 = -142 + 0.1v_1 + 0.2v_2 + 0.4v_3 + 0.1v_4\\
g + v_2 = 271 + 0.1v_1 + 0.2v_2 + 0.2v_3 + 0.2v_4\\
g + v_3 = 91 + 0.3v_1 + 0v_2 + 0.4v_3 + 0.2v_4\\
g + v_4 = 299 + 0.2v_1 + 0.2v_2 + 0.1v_3 + 0.1v_4\\
g = 209 + 0.1v_1 + 0.3v_2 + 0.1v_3 + 0.4v_4
\end{array}$
\end{center}
and have a solution
\begin{center}
$\begin{array}{lcl}
g = 150.78 \\
v_1 = -458.25\\
v_2 = 21.58\\
v_3 = -317.98\\
v_4 = 32.32\\
v_5=0\\
\end{array}$
\end{center}
Using the policy of "advertising on the radio" dealership will have an average of 150.78 units gained per transaction.

Referring to the policy improvement procedure, let us calculate the values of $q_i^k + \sum_{j=1}^N p_{ij}^k v_j$  for all i and k.

% \usepackage{multirow}
\begin{center}
\begin{table}[h]
\centering
\begin{tabular}{|l|l|l|l|}
\hline
i                  & k & \multicolumn{2}{l|}{$q_i^k + \sum_{j=1}^N p_{ij}^k v_j$} \\ \hline
\multirow{5}{*}{1} & 1 & -307.469        &         \\ \cline{2-4} 
                   & 2 & 11.028          &         \\ \cline{2-4} 
                   & 3 & -304.617        &         \\ \cline{2-4} 
                   & 4 & 90.608          &         \\ \cline{2-4} 
                   & 5 & 177.638         & +       \\ \hline
\multirow{5}{*}{2} & 1 & 172.359         &         \\ \cline{2-4} 
                   & 2 & -163.944        &         \\ \cline{2-4} 
                   & 3 & -178.631        &         \\ \cline{2-4} 
                   & 4 & 206.447         & +       \\ \cline{2-4} 
                   & 5 & 189.84          &         \\ \hline
\multirow{5}{*}{3} & 1 & -167.203        &         \\ \cline{2-4} 
                   & 2 & -54.345         &         \\ \cline{2-4} 
                   & 3 & -139.267        &         \\ \cline{2-4} 
                   & 4 & -96.338         &         \\ \cline{2-4} 
                   & 5 & -29.17          & +       \\ \hline
\multirow{5}{*}{4} & 1 & -23.013         &         \\ \cline{2-4} 
                   & 2 & -562.917        &         \\ \cline{2-4} 
                   & 3 & -408.7958       &         \\ \cline{2-4} 
                   & 4 & 74.347          & +       \\ \cline{2-4} 
                   & 5 & -225.7162       &         \\ \hline
\multirow{5}{*}{5} & 1 & 150.779         &         \\ \cline{2-4} 
                   & 2 & 249.14          &         \\ \cline{2-4} 
                   & 3 & 111.984         &         \\ \cline{2-4} 
                   & 4 & 470.231         & +       \\ \cline{2-4} 
                   & 5 & 397.464         &         \\ \hline
\end{tabular}
\end{table}
\end{center}

We see that for i = 1 the value in the right column is maximum for k = 5. Similarly for i = 3. For i = 2, i = 4 and i = 5 the value is maximum for k = 4. In other words, the new policy will be determined by vector D.

\begin{center}
$D = \begin{bmatrix}5\\4\\5\\4\\4\end{bmatrix}$
\end{center}

This means that in the first, third and fifth cities it is most advantageous to use the 5th policy and for the second and fourth cities to use the fourth.

Now we have
\begin{center}
$P = \begin{bmatrix}0.2&0.1&0.1&0.4&0.2\\0.3&0.2&0.3&0&0.2\\0.4&0&0.1&0.4&0.1\\0.3&0.1&0.1&0.5&0\\0.1&0.1&0.1&0.3&0.4\end{bmatrix} \begin{bmatrix}q_i\end{bmatrix} = \begin{bmatrix}209\\330\\119\\536\\674\end{bmatrix}$
\end{center}
Let us solve the equations
\begin{center}
$\begin{array}{lcl}
g +v_1 = 209+0.2v_1+0.1v_2+0.1v_3+0.4v_4+0.2v_5\\
g+v_2 =330+0.3v_1+0.2v_2+0.3v_3+0v_4+0.2v_5\\
g+v_3=119+0.4v_1+0v_2+0.1v_3+0.4v_4+0.1v_5\\
g+v_4=536+0.3v_1+0.1v_2+0.1v_3+0.5v_4+0v_5\\
g+v_5=674+0.1v_1+0.1v_2+0.1v_3+0.3v_4+0.4v_5
\end{array}$
\end{center}
Assuming again that $v_5$ is zero, we get
\begin{center}
$\begin{array}{lcl}
g = 400.633 \\
v_1 = -551.143\\
v_2 = -555.888\\
v_3 = -695.783\\
v_4 = -310.286\\
v_5=0\\
\end{array}$
\end{center}

Note that g has increased from 150.79 to 400.633, so the agency earns an average of 400.633 units per transaction. Using the policy improvement algorithm for these values, we calculate the values $q_i^k + \sum_{j=1}^N p_{ij}^k v_j$ for all i and k.

\begin{center}
\begin{table}[H]
\centering
\begin{tabular}{|l|l|l|l|}
\hline
i                  & k & \multicolumn{2}{l|}{temp} \\ \hline
\multirow{5}{*}{1} & 1 & -617.6337       &         \\ \cline{2-4} 
                   & 2 & -332.1275       &         \\ \cline{2-4} 
                   & 3 & -427.3889       &         \\ \cline{2-4} 
                   & 4 & -164.0598       &         \\ \cline{2-4} 
                   & 5 & -73.5101        & +       \\ \hline
\multirow{5}{*}{2} & 1 & -96.5057        &         \\ \cline{2-4} 
                   & 2 & -362.2529       &         \\ \cline{2-4} 
                   & 3 & -552.2149       &         \\ \cline{2-4} 
                   & 4 & -50.2554        & +       \\ \cline{2-4} 
                   & 5 & -126.8591       &         \\ \hline
\multirow{5}{*}{3} & 1 & -414.7133       &         \\ \cline{2-4} 
                   & 2 & -284.8063       &         \\ \cline{2-4} 
                   & 3 & -487.212        &         \\ \cline{2-4} 
                   & 4 & -407.9878       &         \\ \cline{2-4} 
                   & 5 & -241.1499       & +       \\ \hline
\multirow{5}{*}{4} & 1 & 183.1           &         \\ \cline{2-4} 
                   & 2 & -224.046        &         \\ \cline{2-4} 
                   & 3 & -198.301        &         \\ \cline{2-4} 
                   & 4 & 369.045         & +       \\ \cline{2-4} 
                   & 5 & 47.433          &         \\ \hline
\multirow{5}{*}{5} & 1 & -206.5734       &         \\ \cline{2-4} 
                   & 2 & 46.5366         &         \\ \cline{2-4} 
                   & 3 & -223.3673       &         \\ \cline{2-4} 
                   & 4 & 262.6328        & +       \\ \cline{2-4} 
                   & 5 & 189.2191        &         \\ \hline
\end{tabular}
\end{table}
\end{center}

The new policy is thus determined by the vector D.

\begin{center}
$D = \begin{bmatrix}5\\4\\5\\4\\4\end{bmatrix}$
\end{center}

Which coincides with the vector of the previous policy, the process has converged, and the value reached its maximum equal to 400.633. The agency must actively search for clients in the first and third city, but use TV adverts in the rest. The application of this policy will give, on average, per transaction, an income of 419.2 units.

\end{document}